\documentclass[twocolumn,notitlepage,prb,color,superscriptaddress,psfig,showpacs,amsmath,amssymb,nobibnotes,longbibliography]{revtex4-1}

\setcitestyle{square,numbers}
\usepackage{graphicx}
\usepackage{float}
\usepackage{epstopdf}
\usepackage{dcolumn}
\usepackage{hyperref}
\hypersetup{
    colorlinks   = true, 
    urlcolor     = blue, 
    linkcolor    = blue, 
    citecolor    = blue 
}
\usepackage{color}
\usepackage{amsmath}
\usepackage{xspace}

\newcommand{\CCL}{CrCl$_3$\xspace}
\newcommand{\CGT}{Cr$_2$Ge$_2$Te$_6$\xspace}

\begin{document}
	
\title{Electron spin resonance and ferromagnetic resonance spectroscopy\\ in the high-field phase of the van der Waals magnet \CCL}

\author{J.~Zeisner}
\thanks{These authors contributed equally to this work.}
\affiliation{Leibniz IFW Dresden, D-01069 Dresden, Germany}
\affiliation{Institute for Solid State and Materials Physics, TU Dresden, D-01062 Dresden, Germany}
\author{K.~Mehlawat}
\thanks{These authors contributed equally to this work.}
\affiliation{Leibniz IFW Dresden, D-01069 Dresden, Germany}
\affiliation{W{\"u}rzburg-Dresden Cluster of Excellence ct.qmat, Germany}
\author{A.~Alfonsov}
\affiliation{Leibniz IFW Dresden, D-01069 Dresden, Germany}
\author{M.~Roslova}
\affiliation{Department of Materials and Environmental Chemistry, Stockholm University, SE-106 91, Stockholm, Sweden}
\author{T.~Doert}
\affiliation{Faculty of Chemistry and Food Chemistry, TU Dresden, D-01062 Dresden, Germany}
\author{A.~Isaeva}
\affiliation{Leibniz IFW Dresden, D-01069 Dresden, Germany}
\affiliation{Institute for Solid State and Materials Physics, TU Dresden, D-01062 Dresden, Germany}
\affiliation{W{\"u}rzburg-Dresden Cluster of Excellence ct.qmat, Germany}
\author{B.~B{\"u}chner}
\affiliation{Leibniz IFW Dresden, D-01069 Dresden, Germany}
\affiliation{Institute for Solid State and Materials Physics, TU Dresden, D-01062 Dresden, Germany}
\affiliation{W{\"u}rzburg-Dresden Cluster of Excellence ct.qmat, Germany}
\author{V.~Kataev}
\affiliation{Leibniz IFW Dresden, D-01069 Dresden, Germany}
\date{\today}

\begin{abstract}
We report a comprehensive high-field/high-frequency electron spin resonance (ESR) study on single crystals of the van der Waals magnet \CCL. This material, although being known for quite a while, has received recent significant attention in a context of the use of van der Waals magnets in novel spintronic devices. Temperature-dependent measurements of the resonance fields were performed between 4 and 175\,K and with the external magnetic field applied parallel and perpendicular to the honeycomb planes of the crystal structure. These investigations reveal that the resonance line shifts from the paramagnetic resonance position already at temperatures well above the transition into a magnetically ordered state. Thereby the existence of ferromagnetic short-range correlations above the transition is established and the intrinsically two-dimensional nature of the magnetism in the title compound is proven. To study details of the magnetic anisotropies in the field-induced effectively ferromagnetic state at low temperatures, frequency-dependent ferromagnetic resonance (FMR) measurements were conducted at 4\,K. The observed anisotropy between the two magnetic-field orientations is analyzed by means of numerical simulations based on a phenomenological theory of FMR. These simulations are in excellent agreement with measured data if the shape anisotropy of the studied crystal is taken into account, while the magnetocrystalline anisotropy is found to be negligible in \CCL. The absence of a significant intrinsic anisotropy thus renders this material as a practically ideal isotropic Heisenberg magnet.
\end{abstract}

\maketitle

\section{Introduction}
\label{sec:introduction}
\begin{figure*}
	\includegraphics[width=0.4\textwidth,angle=-90]{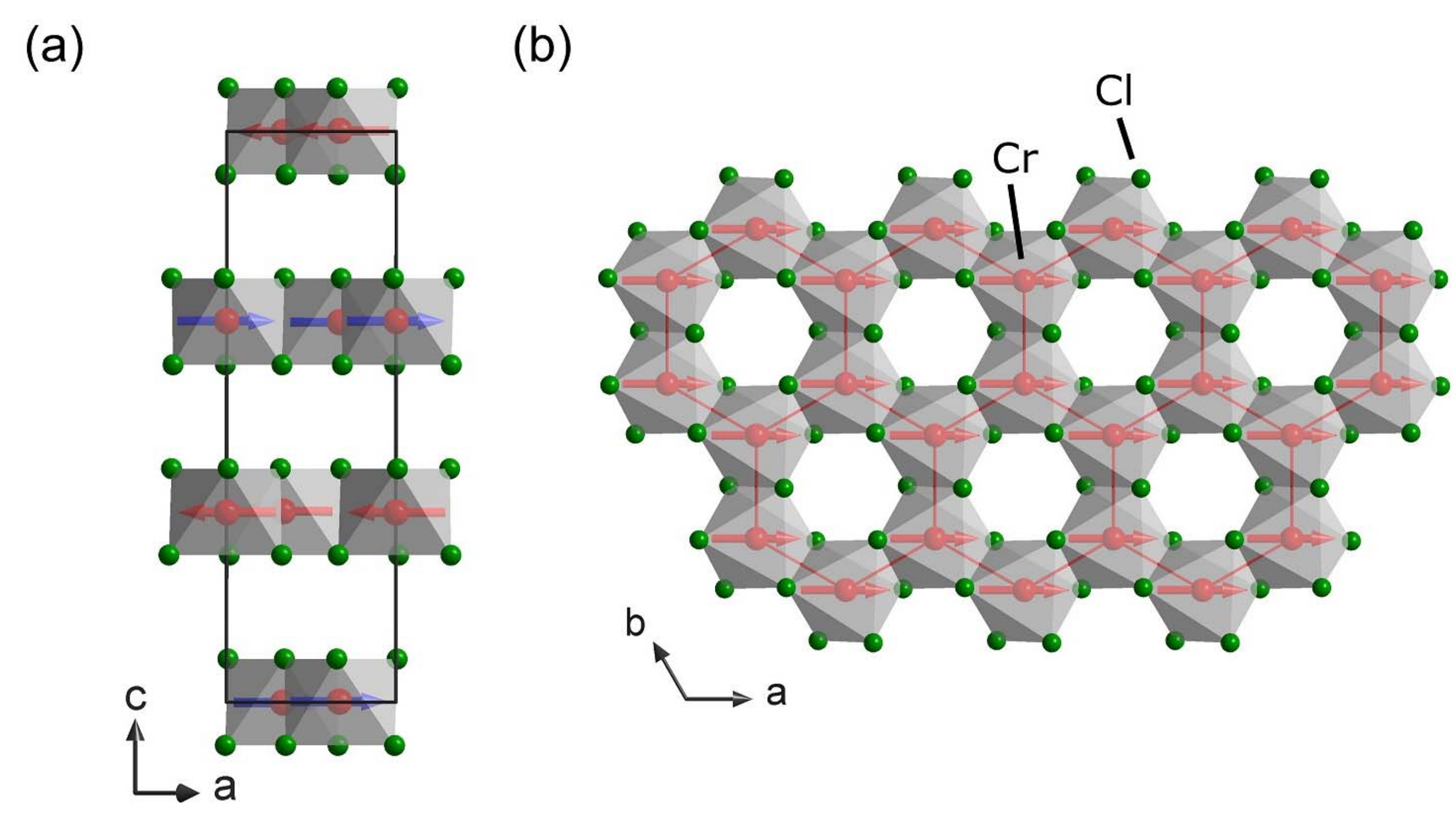}
	\caption{Crystal and magnetic structures of \CCL at low temperatures. (a) View along the $b$-axis of the unit cell (non-primitive hexagonal) of the rhombohedral low-temperature structure (space group $R\overline{3}$ \cite{Morosin1964}). The magnetic Cr$^{3+}$ ions (red spheres) are octahedrally coordinated by the Cl ligands (green spheres). (b) These CrCl$_6$ octahedra build an edge-sharing network in the $ab$ plane which effectively leads to a honeycomb-like arrangement of the magnetic ions (illustrated by solid red lines). Each of these honeycomb layers in the $ab$ plane is well separated from its neighbors along the $c$ axis by the van der Waals gap, as shown in (a). At temperatures below the transition into a magnetically long-range ordered state, spins in the honeycomb layers are coupled ferromagnetically and oriented within the $ab$-plane, while spins in neighboring layers are coupled by a weaker antiferromagnetic interaction \cite{Cable1961,McGuire2017}. The resulting spin structure in the magnetically ordered state at zero magnetic field \cite{Cable1961,McGuire2017} is schematically illustrated by the arrows at the Cr sites. Crystallographic data are taken from Ref.~\cite{Morosin1964}.}
	\label{fig:structure}
\end{figure*}

Magnetic van der Waals materials belong to a class of physical systems that currently receive considerable attention in solid state and materials research \cite{Park2016,Gong2017,Huang2017,Burch2018,Gibertini2019,Gong2019}. As a common feature these materials crystallize in a layered structure with the individual layers being separated by the so-called van der Waals gap, see Fig.~\ref{fig:structure}. The weak van der Waals coupling between the adjacent layers results in dominating magnetic interactions within the layers while magnetic couplings between the layers remain relatively weak. Thus, van der Waals magnets can be considered as quasi-two-dimensional magnetic systems. These systems enable experimental studies of the specific magnetic properties arising from the interplay of an effectively reduced dimensionality and the respective single-ion anisotropies determined by the type of magnetic ions and their local environments, see for instance \cite{Burch2018}. Moreover, the weak van der Waals bonds between the layers allow mechanical exfoliation of bulk crystals down to the few-layer or even monolayer limit \cite{Gong2017,Huang2017}, thereby approaching the experimental systems to the true two dimensional (2D) limit. In addition, the combination of various materials sharing similar layer structures and the weak van der Waals couplings between the layers open up numerous possibilities for the creation of (magnetic) van der Waals heterostructures \cite{Geim2013,Gong2019,Gibertini2019}. These stacks of several few-layer crystals of different materials are a promising route towards electronic devices with specifically tailored magnetic properties, see, e.g., Ref.~\cite{Gong2019} and references therein. In both respects -- the fundamental study of 2D magnetism and its application in the framework of magnetic heterostructures -- the characterization of magnetic anisotropies in van der Waals materials represents a key task. Magnetic resonance spectroscopies are valuable tools to accomplish this task due to their high sensitivity with respect to the presence of magnetic anisotropies in a system. As an example, in a previous work \cite{Zeisner2019CGT} some of the authors  quantitatively investigated the magnetic anisotropies in the van der Waals magnet \CGT by means of electron spin resonance (ESR) and ferromagnetic resonance (FMR) studies. Here, we report a characterization of the (effective) magnetic anisotropy in the high-field phase of the related compound \CCL. The last is a member of the family of transition-metal trihalides whose magnetic ions (here: Cr$^{3+}$, $3d^3$, $S=3/2$, $L=3$) are situated in the center of an octahedron built by the halogen ligands (here: Cl$^-$ ions). These octahedra form edge-sharing networks in the crystallographic $ab$ plane which effectively results in a magnetic honeycomb lattice, see Fig.~\ref{fig:structure}.

It is worthwhile to mention that the title compound belongs to the first materials studied using the ESR technique by E. K. Zavoisky, the pioneer of this spectroscopy (see, for instance, \cite{Zavoisky1946,Kochelaev1995}). However, in the context of magnetic van der Waals materials the magnetic properties of \CCL came back to the focus of current research interest \cite{Glamazda2017,McGuire2017,Besbes2019,Bastien2019,Abramchuk2018,Zhang2015,Webster2018,Cai2019,Bykovetz2019,Klein2019,MacNeill2019}. Basic magnetic properties of bulk \CCL crystals have been reported in the past decades, see, e.g., Refs.~\cite{Hansen1958,Hansen1959,Bizette1961,Narath1965,Kuhlow1982,McGuire2017,Bastien2019}. In particular, a two-step transition into a magnetically long-range ordered state was observed \cite{Kuhlow1982,McGuire2017,Bastien2019}. Below a temperature $T_c^{\text{2D}} \sim 17$\,K, spins within the honeycomb layers order ferromagnetically and align parallel to the honeycomb plane, i.e., the $ab$ plane [cf. Fig.~\ref{fig:structure}(b)], while spins of individual Cr layers are not coupled to spins in the neighboring layers \cite{Kuhlow1982,McGuire2017,Bastien2019}. Consequently, the ordering at around 17\,K is of an effectively 2D nature. The ferromagnetic character of the dominant exchange interactions between spins in the $ab$ plane is also evidenced by the positive Curie-Weiss temperatures $\Theta_{\text{CW}}$ between 27 and 43\,K derived from measurements of the static susceptibility \cite{McGuire2017,Roslova2019,Starr1940,Hansen1959,Bastien2019,Abramchuk2018}. Below a temperature $T_N^{\text{3D}}$ of about 14\,K \cite{McGuire2017,Bastien2019} (15.5\,K in Ref.~\cite{Kuhlow1982}) \CCL enters into a three-dimensional (3D) antiferromagnetically ordered state at zero magnetic field. In this magnetic phase, ferromagnetically ordered spins in the honeycomb layers are coupled by antiferromagnetic interactions between neighboring layers \cite{Cable1961}, as illustrated in Fig.~\ref{fig:structure}(a). However, the long range antiferromagnetic order can be suppressed already in relatively small magnetic fields of about 0.6\,T  (external field applied perpendicular to the honeycomb planes, i.e., $H \parallel c$) and 0.25\,T (external field applied in the honeycomb planes, i.e., $H \perp c$) at 2\,K \cite{McGuire2017}, respectively. Above these fields, a saturation of the magnetization at around 3\,$\mu_B$/Cr was observed \cite{McGuire2017,Bastien2019}. The low values of the saturation fields thus confirm the (relative) weakness of the antiferromagnetic interlayer couplings. Moreover, if demagnetization effects are taken into account, the saturation fields become almost identical for both orientations of the external field with respect to the honeycomb layers \cite{McGuire2017,Bastien2019}. Therefore, the experimentally observed magnetic anisotropies appear to be dominated by dipole-dipole interactions which are at the origin of demagnetization fields and the so-called shape anisotropy \cite{McGuire2017,Bastien2019}. The apparent size of the magnetic anisotropy thus depends strongly on the dimensions of the studied samples whereas the intrinsic magnetocrystalline is expected to be much weaker, if it exists at all. In order to disentangle and quantify the two contributions to an anisotropic magnetic response, which could be of relevance in the case of \CCL, we studied in this work the details of the magnetic anisotropies in the field-induced ferromagnetic-like phase of the title compound by means of high-field/high-frequency (HF) FMR over a wide range of frequencies at magnetic fields exceeding the low-temperature saturation fields of \CCL.

\section{Samples and experimental methods}
\label{sec:methods}

Single-crystalline \CCL samples used in this study were synthesized by a chemical vapor transport reaction between Cr metal and Cl$_2$ gas as described in detail in Ref.~\cite{Roslova2019}. The results of single-crystal X-ray diffraction studies at room temperature confirming the monoclinic (space group $C2/m$) high-temperature modification of the title compound are also reported there. Moreover, magnetic properties of the samples were studied in Ref.~\cite{Bastien2019} by means of specific heat as well as static and dynamic magnetization measurements. These are consistent with the findings reported in previous studies \cite{McGuire2017,Kuhlow1982,Bizette1961}, in particular, they confirm the presence of two successive magnetic phase transitions mentioned in the previous section. This does not only demonstrate the high quality of the used \CCL crystals but also ensures the comparability of the magnetic properties reported in this work and in the literature \cite{McGuire2017,Kuhlow1982,Bizette1961,MacNeill2019}. Additional characterization details can be found in the Appendix.

The compound \CCL is known to undergo a structural phase transition at temperatures around 240\,K from the high-temperature monoclinic phase to an rhombohedral phase (space group $R\overline{3}$) at lower temperatures \cite{Morosin1964,McGuire2017}. These two structural modifications differ mainly in the stacking sequence of the honeycomb layers while the structure within the layers is very similar in both phases \cite{McGuire2017}. The crystal structure of the rhombohedral modification is shown in Fig.~\ref{fig:structure} as the focus of the present study lies on the magnetic properties at lower temperatures. In this phase, individual honeycomb layers are stacked along the $c$ axis in an -ABC-sequence. Consequently, each unit cell contains three layers along the $c$ axis, see Fig.~\ref{fig:structure}(a). The strong chemical bonding within the $ab$ plane and the comparatively weak van der Waals couplings between the layers result in flat, platelet-like single crystals allowing an easy identification of the $c$ axis as the direction perpendicular to the platelet plane.

The ESR and FMR measurements were carried out using two different setups. For continuous wave (cw) HF-ESR/HF-FMR studies a homemade spectrometer was employed. The spectrometer consists of a network vector analyzer (PNA-X from Keysight Technologies) for generation and detection of microwaves in the frequency range from 20 to 330\,GHz, oversized waveguides, and a superconducting solenoid (Oxford Instruments) providing magnetic fields up to 16\,T. The magnetocryostat is equipped with a variable temperature insert that enables measurements in the temperature range between 1.8 and 300\,K. All HF measurements presented in the following section were carried out in transmission geometry employing the Faraday configuration. In addition, cw ESR/FMR measurements at a fixed microwave frequency of about 9.6\,GHz and in magnetic fields up to 0.9\,T were performed using a commercial X-band spectrometer (EMX from Bruker). This spectrometer is equipped with a helium flow cryostat (ESR900 from Oxford Instruments) and a goniometer, allowing temperature-dependent measurements in the range 4 - 300\,K and angle-dependent studies, respectively.

\section{Results and discussion}
\label{sec:results}

In the following, results of systematic ESR measurements are presented and discussed. These measurements were carried out with the external magnetic field $H$ applied parallel and perpendicular to the crystallographic $c$ axis, respectively. Since the focus of the present study lies on a detailed investigation of the (effective) magnetic anisotropies in the field-polarized ferromagnetic state of \CCL at low temperatures, this work is mainly concerned with the behavior of the resonance field $H_{\text{res}}$ as a function of temperature and microwave frequency over a broad frequency range. Further aspects of the magnetic properties of \CCL derived from ESR measurements at lower microwave frequencies, such as the excitations of the 3D antiferromagnetic state and the spin dynamics above the magnetic phase transitions, were reported in Refs.~\cite{Chehab1991,MacNeill2019}.

\subsection{Temperature dependence}
\label{subsec:Temp-dep}

\begin{figure}
	\includegraphics[width=\columnwidth]{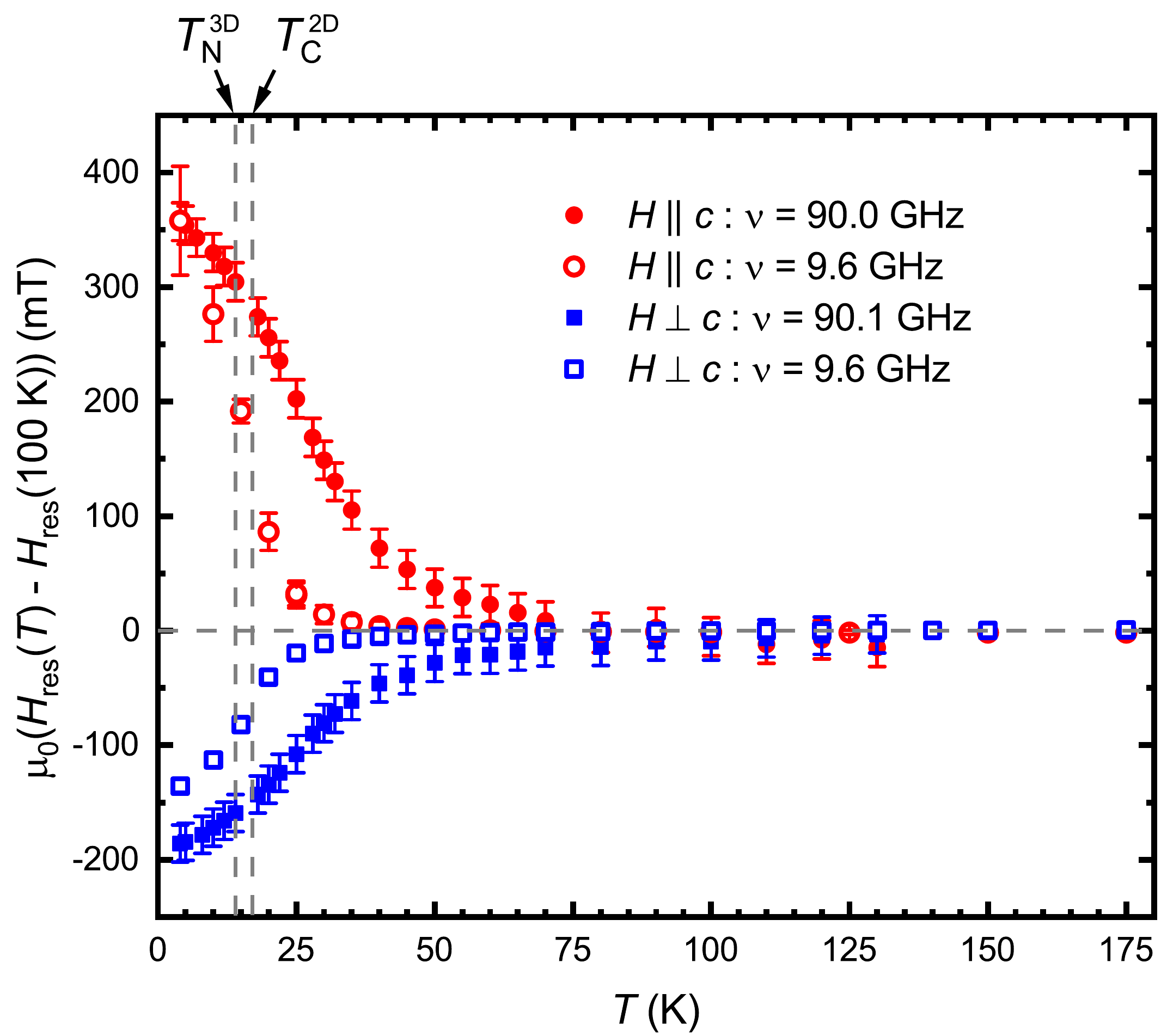}
	\caption{Temperature dependence of the resonance shift $\delta H(T) = H_{\text{res}}(T)-H_{\text{res}}(100\,K)$ at microwave frequencies of about 9.6 and 90\,GHz (open and filled symbols, respectively). In these measurements the external magnetic field was oriented parallel (red circles) as well as perpendicular (blue squares) to the $c$ axis. The dashed horizontal line represents the zero shift $\delta H = 0$ expected for an uncorrelated paramagnet. Vertical dashed lines indicate the zero-field transition temperatures $T_c^{\text{2D}}$ and $T_N^{\text{3D}}$ of the transition into the 2D ferromagnetically ordered phase and the 3D antiferromagnetically ordered state \cite{McGuire2017}, respectively.}
	\label{fig:temperature_dependence}
\end{figure}

The temperature dependence of the resonance shift $\delta H(T) = H_{\text{res}}(T)-H_{\text{res}}(100\,K)$ measured at low and high microwave frequencies $\nu$ and in both magnetic field configurations is shown in Fig.~\ref{fig:temperature_dependence}. This quantity is a measure of the deviation of the resonance field at a given temperature $T$ from the ideal paramagnetic resonance position. This position is determined by the standard resonance condition of a paramagnet \cite{AbragamBleaney}
\begin{equation}
\label{eq:EPR}
\nu = g\mu_B\mu_0H_{\text{res}}/h \ \ \ .
\end{equation}
Here, $g$ denotes the $g$ factor of the resonating spins and $\mu_B$, $\mu_0$, and $h$ are Bohr's magneton, the vacuum permeability, and Planck's constant, respectively. In the present case, the resonant shift was determined with respect to the expected resonance field at 100\,K which was calculated according to Eq.~(\ref{eq:EPR}) using the $g$ factor derived from frequency-dependent measurements at 100\,K, see below. Upon lowering the temperature below $\sim$75\,K, the resonance position is shifted progressively to smaller fields when the external magnetic field is applied perpendicular to the $c$ axis, resulting in a negative shift $\delta H$. For $H \parallel c$ this trend is reversed yielding a positive resonance shift. Thus, based on the qualitative behavior of the temperature-dependent resonance shift, it can be concluded that the experimentally observable (effective) magnetic easy direction lies within the $ab$ plane which is in agreement with the proposed spin structure \cite{Cable1961} and previous magnetization measurements \cite{Bizette1961,McGuire2017,Bastien2019}. Moreover, the $\delta H(T)$ curves obtained for $H \parallel c$ and $H \perp c$ are asymmetric with respect to the $\delta H = 0$ line of an ideal paramagnet, see Fig.~\ref{fig:temperature_dependence}. This asymmetry is a direct consequence of the different frequency-field dependencies $\nu(H_{\text{res}})$ expected for the two different field orientations in a ferromagnetically ordered system \cite{Turov,Farle1998}. In particular, a shift of the resonance line should be larger when the external magnetic field is oriented parallel to the magnetic anisotropy axis which in the case of \CCL is the magnetic hard axis parallel to the crystallographic $c$~axis. The onset of a finite resonance shift already at temperatures significantly above the ordering temperature $T_c^{\text{2D}}$ provides clear evidence for the low-dimensional character of the spin correlations in this material as it was discussed in the context of one-dimensional systems, for instance in Refs.~\cite{Nagata1972,Okuda1972,Oshima1976}. Moreover, the 2D nature of magnetic correlations above $T_c^{\text{2D}}$ was reported in Ref.~\cite{Chehab1991} based on  measurements of the resonance shift and the linewidth angular dependence at various temperatures and at low microwave frequencies of about 9.4\,GHz. While our measurements of $\delta H(T)$ at 9.6\,GHz are consistent with this previous study \cite{Chehab1991}, we observed the onset of a finite resonance shift at higher temperatures when employing frequencies of about 90\,GHz. Thus, the higher external magnetic fields associated with the higher microwave frequencies strengthen the ferromagnetic correlations responsible for the shift of the resonance line, similar to the situation found in the related van der Waals magnet \CGT \cite{Zeisner2019CGT}. Finally, it is worthwhile mentioning that the appearance of short-range correlations at temperatures far above the magnetic ordering temperature is consistent with the deviation of the temperature-dependent static susceptibility from a Curie-Weiss behavior already below $\sim$125\,K, which was reported in Ref.~\cite{Roslova2019}. Taken together, temperature-dependent measurements of the resonance shift at two different frequencies and field orientations demonstrate, first, the apparent easy-plane type magnetic anisotropy and, second, the 2D character of dynamic spin correlations in \CCL far above the long-range ordering temperatures.

\subsection{Frequency dependence}
\label{subsec:freq-dep}

\begin{figure*}
	\includegraphics[width=\textwidth]{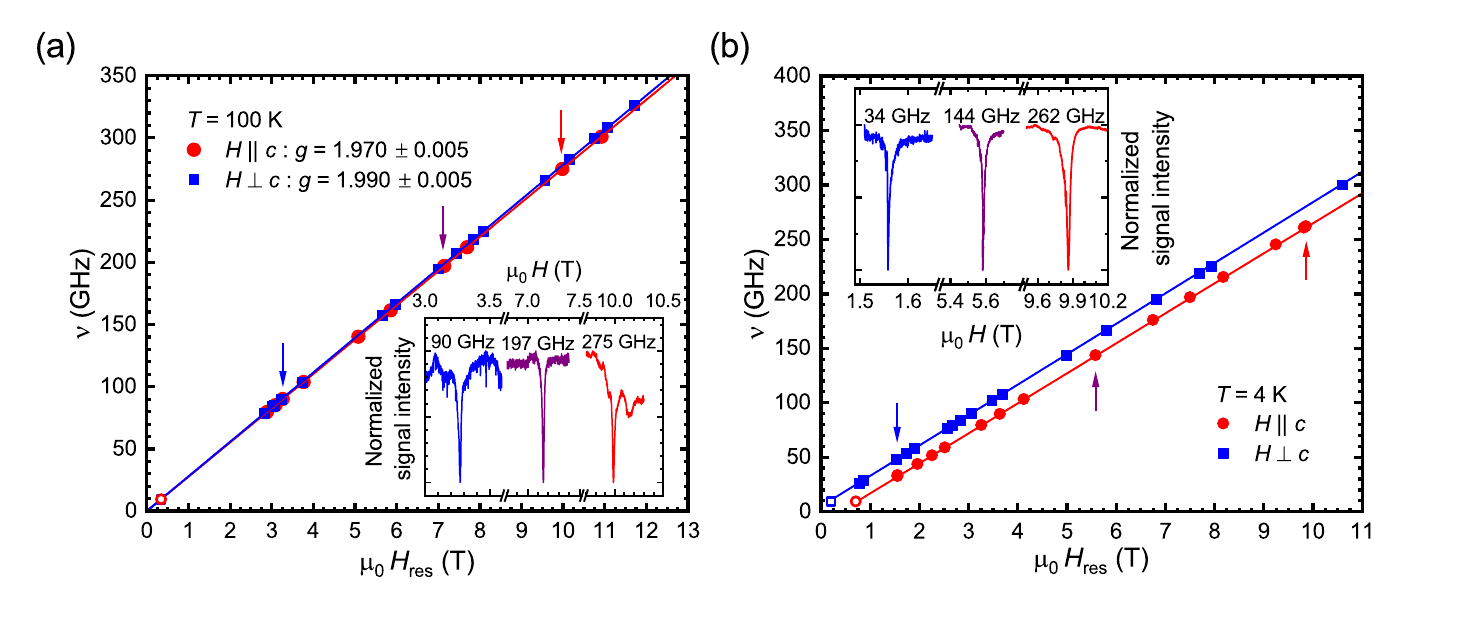}
	\caption{Frequency dependence of the resonance field $H_{\text{res}}$ at 100\,K (a) and 4\,K (b), respectively. To quantitatively investigate the magnetic anisotropies in \CCL, measurements were carried out with the external magnetic field applied parallel (red circles) and perpendicular (blue squares) to the crystallographic $c$ axis, respectively. Open symbols correspond to the resonance fields measured at $\nu = 9.6$\,GHz. Solid lines in (a) are linear fits to the $\nu(H_{\text{res}})$ dependencies according to the standard resonance condition of a paramagnet given in Eq.~(\ref{eq:EPR}) \cite{AbragamBleaney}. At 4\,K, the measured frequency-field dependencies were simulated using Eq.~(\ref{eq:FMR_resonance_frequency}) which describes the resonance frequency in the case of FMR. Corresponding simulations are shown in (b) as solid lines. Exemplary spectra recorded with $H \parallel c$ are presented for both temperatures in the respective insets. For comparison, all spectra shown were normalized (note the different breaks in the field axes). Arrows in the $\nu(H_{\text{res}})$ denote the positions of the spectra given in the insets.}
	\label{fig:frequency_dependence}
\end{figure*}

To shed light on the details of the magnetic anisotropies, frequency-dependent investigations were conducted at 4 and 100\,K, i.e., at temperatures deep in the magnetically ordered state and well above the magnetic phase transition, respectively. Exemplary spectra are presented in the insets of Fig.~\ref{fig:frequency_dependence}. At both temperatures, ESR/FMR spectra consist of a single narrow resonance line with typical linewidths (full width at half maximum) of about 25\,mT at 100\,K. The small linewidths allow an easy and precise determination of the resonance fields which correspond to the positions of the minima in the microwave transmission. The resulting frequency-field diagrams are shown in the main panels of Fig.~\ref{fig:frequency_dependence}. At 100\,K a linear frequency-field dependence $\nu (H_{\text{res}})$ is observed for both orientations of the external magnetic field. Such a behavior is expected in the paramagnetic regime of \CCL and can be well described by the standard resonance condition of a paramagnet [Eq.~(\ref{eq:EPR})] \cite{AbragamBleaney}. Fits to the data according to Eq.~(\ref{eq:EPR}) are shown in Fig.~\ref{fig:frequency_dependence}(a) as solid lines and yielded $g$ factors $g_{\parallel} = 1.970 \pm 0.005$ and $g_{\perp} = 1.990 \pm 0.005$ for $H \parallel c$ and $H \perp c$, respectively. The experimentally determined $g$ factors are in good agreement with the values anticipated for Cr$^{3+}$ ions (3$d^3$, $S = 3/2$, $L = 3$) in an octahedral crystal field \cite{AbragamBleaney}. The merely small deviation of $g$ from the free-electron $g$ factor of 2 as well as the slight anisotropy observed in the measurements indicate that the magnetism in \CCL is largely dominated by the spin degrees of freedom while the orbital angular momentum is practically completely quenched in the first order. The mentioned small deviations from the ideal spin-only behavior result, most likely, from second-order spin-orbit coupling effects \cite{AbragamBleaney}. The $g$ factors derived in this work from frequency-dependent measurements are, moreover, consistent with the saturation magnetization of about 3\,$\mu_B$/Cr which was experimentally observed \cite{McGuire2017,Bastien2019} and is theoretically expected for spins $S = 3/2$ and a $g$ factor of 2. These observations already suggest that spin-orbit coupling and, consequently, the intrinsic magnetocrystalline anisotropies are very weak (or even negligible) in \CCL, as it was also mentioned in the literature \cite{Bizette1961,McGuire2017,Bastien2019}.

For a detailed analysis of the relevant anisotropies present in the title compound, frequency-dependent measurements were conducted at 4\,K, i.e., in the magnetically ordered state. We emphasize that the lowest frequency used in the systematic HF measurements corresponds to a resonance field of about 0.75\,T [cf. Fig.~\ref{fig:frequency_dependence}(b)] which exceeds the reported saturation fields \cite{Bizette1961}. Therefore, the field-polarized, effectively 2D ferromagnetic state of \CCL is probed by our HF magnetic resonance measurements allowing us to describe the obtained $\nu (H_{\text{res}})$ dependencies in the framework of a theory of FMR that is applicable in a single-domain ferromagnetic material, as will be discussed in the following section. In contrast to the measurements in the paramagnetic state, the 4\,K data show a clear anisotropy regarding the two magnetic-field orientations which is consistent with the resonance shifts of opposite sign observed in the temperature-dependent studies (see Fig.~\ref{fig:temperature_dependence}). If the external magnetic field is applied within the $ab$ plane ($H \perp c$), the resonance positions are systematically shifted towards smaller magnetic fields (with respect to the paramagnetic resonance positions) for all measured frequencies. For $H \parallel c$ a negative intercept with the frequency axis of the $\nu (H_{\text{res}})$ diagram is observed due to the shift of the resonance positions to higher magnetic fields.  Qualitatively, such a behavior is expected in the case of a ferromagnetically ordered system with an easy-plane anisotropy \cite{Farle1998,Skrotskii1966}. In a semi-classical picture, these shifts of the resonance position result from anisotropic internal fields in the magnetic crystal. These fields are caused, in turn, by dipole-dipole interactions between the magnetic moments associated with the spins or an intrinsic magnetocrystalline anisotropy, i.e., by spin-orbit coupling. In the following section the resonance fields in the ferromagnetic state will be simulated employing a theory based on this semi-classical description of magnetic resonance. This allows to determine quantitatively the contributions of the two possible sources of magnetic anisotropy in \CCL.    
	 
\subsection{Analysis of the magnetic anisotropies}
\label{subsec:analysis}

As mentioned in the previous sections, it is possible to describe the experimentally observed frequency-field dependence in the ferromagnetic state at 4\,K in the framework of a semi-classical, phenomenological theory of FMR \cite{Skrotskii1966,Smit1955,Farle1998}. Conceptually, the difference between ESR and FMR lies in the fact, that ESR is the resonant excitation of individual paramagnetic spins within a magnetic system, while FMR describes the resonance of the total magnetization $\boldsymbol{M}$ in a ferromagnetically ordered material. Thus, the resonance frequencies expected for a correlated ferromagnetic spin system can be calculated by considering the classical vector of the macroscopic magnetization and the appropriate free energy density $F$ \cite{Skrotskii1966,Smit1955,Farle1998}. In this case, the resonance frequency is given by the following expression \cite{Skrotskii1966,Smit1955,Farle1998}
\begin{equation}
\label{eq:FMR_resonance_frequency}
\nu_{res}^2 = \frac{g^2 \mu_B^2}{h^2 M_s^2\sin^2\theta} \bigg(\frac{\partial^2{F}}{\partial \theta ^2}\frac{\partial^2{F}}{\partial \varphi^2} - \Big(\frac{\partial^2{F}}{\partial \theta \partial \varphi}\Big)^2\bigg) \ \ \  ,
\end{equation}
where $M_s$ is the saturation magnetization and $\varphi$ and $\theta$ denote the spherical coordinates of the magnetization vector $\boldsymbol{M}(M_s,\varphi,\theta)$. Note that in the present case the spherical coordinate system is chosen such that the $z$ axis coincides with the crystallographic $c$ axis in the low-temperature structure of \CCL. For a calculation of the resonance position under given experimental conditions, i.e., a specific orientation of the external magnetic field relative to the studied sample and a fixed microwave frequency, Eq.~(\ref{eq:FMR_resonance_frequency}) has to be evaluated at the equilibrium position $(\varphi_0,\theta_0)$ of the macroscopic magnetization vector. The orientation of $\boldsymbol{M}$ in the equilibrium state is found numerically by minimizing $F$ with respect to the spherical coordinates, taking into account the particular experimental conditions. Since the qualitative considerations regarding the temperature dependence of the resonance shift as well as the $\nu (H_{\text{res}})$ dependencies at 4\,K suggest an easy-plane type anisotropy, a uniaxial magnetocrystalline anisotropy was included in the free energy density term as an initial approach to describe the anisotropies in \CCL. The free energy density is then given (in SI units) by 
\begin{equation} 
\label{eq:free_energy_density}
\begin{split}
F &= -\mu_0\boldsymbol{H}\cdot\boldsymbol{M} - K_U\cos^2(\theta) + \\ & \ \ \ \  \frac{1}{2}\mu_0 M^2_s(N_x\sin^2(\theta)\cos^2(\varphi) + \\ & \ \ \ \  N_y\sin^2(\theta)\sin^2(\varphi) + N_z\cos^2(\theta)) \ \ \ .
\end{split}
\end{equation}
The first term is the Zeeman-energy density describing the coupling between the magnetization vector $\boldsymbol{M}$ and the external magnetic field $\boldsymbol{H}$. The second term represents the uniaxial magnetocrystalline anisotropy whose strength is parametrized by the energy density $K_U$. Finally, the third contribution to $F$ is the shape anisotropy energy density which is characterized by the demagnetization factors $N_x$, $N_y$, and $N_z$ \cite{Osborn1945,Cronemeyer1991}. These factors are determined by the dimensions of the crystals under study. In the present case, the shape of the measured platelet-like \CCL crystal was described by the demagnetization factors of an extended flat plate ($N_x = N_y = 0, N_z = 1$ \cite{Blundell2001}), whose $xy$ plane corresponds to the crystallographic $ab$ plane and the $z$ axis lies parallel to the $c$ axis. This approximation to the real sample shape is justified by the platelet-like shape of the studied \CCL crystal whose lateral dimensions in the $ab$ plane (of about 400\,$\mu$m $\times$ 450\,$\mu$m) are much larger than the thickness of the crystal along the $c$ axis. Due to experimental reasons, the sample thickness could not be measured precisely which hampered a determination of the demagnetization factors solely based on the sample dimensions. However, the deviations between the approximated and the true demagnetization factors can be expected to be small. In addition to the sample shape, the value of the saturation magnetization $M_s$ enters into the simulation of the frequency-field dependence. For the simulations presented in the following, the reported saturation magnetization of about 3\,$\mu_B$/Cr at 1.8\,K \cite{Bastien2019} was used which corresponds to $M_s \approx 314.97\times 10^3$\,J/Tm$^3$. Furthermore, the $g$ factors determined independently from the frequency-dependent measurements at 100\,K and for both field orientations (see Sec.~\ref{subsec:freq-dep}) were set as fixed parameters in the simulations. Thus, the only free parameter, which was adjusted to match simulated with measured data, was the magnetocrystalline anisotropy energy density $K_U$. 

The final results of the simulations are presented in Fig.~\ref{fig:frequency_dependence}(b) as solid lines and show an excellent agreement between the simulated and the measured resonance positions. Most importantly, this agreement could be accomplished by solely taking into account the shape anisotropy, i.e., by setting $K_U$ to zero. Thus, it is ultimately proven that the anisotropy observed in dynamic and static magnetic investigations of \CCL \cite{Bizette1961,McGuire2017,Bastien2019,MacNeill2019} is, indeed, due to the shape anisotropy caused by long-range dipole-dipole interactions, whereas the intrinsic magnetocrystalline anisotropy can be neglected. The present work therefore could verify by means of highly sensitive HF magnetic resonance investigations the conclusions of these previous studies \cite{Bizette1961,McGuire2017,Bastien2019,MacNeill2019}  and supports the results of the recent first-principle calculations of the magnetic anisotropy of \CCL monolayers \cite{Xue2019}. It can be concluded that, intrinsically, \CCL is magnetically isotropic while the apparent anisotropy observed in experiments can be tuned (within certain limits) by choosing a particular sample shape. Moreover, \CCL might serve as a valuable reference system for future magnetic resonance studies of other van der Waals magnets, since it illustrates the pure effect of the shape anisotropy on, e.g., the frequency-field dependence. As similar sample shapes can be expected for this large family of layered materials, it follows that it is very important to take into account this source of magnetic anisotropy when aiming at a precise quantification of magnetocrystalline anisotropies in van der Waals magnets. 

\section{Conclusions}
\label{sec:conclusions}

In summary, we studied the details of the magnetic anisotropies in the van der Waals magnet \CCL by means of systematic HF ESR and FMR measurements. By extending the frequency range of previous works \cite{Chehab1991,MacNeill2019}, the field-polarized, effectively ferromagnetic low-temperature state of the title compound was investigated. Numerical simulations of the measured frequency-field dependence at 4\,K revealed that the anisotropy observed experimentally is governed by the shape anisotropy of the studied \CCL crystal. In contrast, the intrinsic magnetocrystalline anisotropy is found to be negligible in this compound, thus supporting the conclusions drawn in previous studies \cite{Bizette1961,McGuire2017,Bastien2019,MacNeill2019}. Considering the large current scientific interest in magnetic van der Waals materials, \CCL may serve as a reference for future quantitative analyses of magnetic anisotropies in related layered magnets, since it provides an excellent example for the impact of the particular sample shape on the apparent magnetic anisotropy. Finally, temperature-dependent measurements of the resonance shift showed the onset of a finite shift and, thus, the existence of short-range spin correlations already at temperatures well above the magnetically ordered state. This observation provides further evidence for the intrinsically 2D nature of the magnetism  in the van der Waals magnet \CCL. 

Considering a continuously increasing number of quasi-2D van der Waals compounds exhibiting a rich variety of intriguing magnetic properties, it is appealing to apply systematically the frequency-tunable high-field ESR spectroscopy to investigate the spin dynamics and in particular the magnetic anisotropy of these materials. The latter appears to be a key factor for determining the type of a magnetically ordered ground state. For example, the members of the family of the van der Waals layered metal phosphorous trichalcogenides  MPS$_3$ (M = Mn, Fe, Ni) \cite{Wang2018} feature different types of magnetic anisotropy \cite{Joy1992} and exhibit dissimilar antiferromagnetically ordered ground states, as, e.g., was illustrated in Ref.~\cite{Chu2020}. Quantification of the parameters of magnetic anisotropy with ESR spectroscopy may shed more light on a possible relationship between the anistropy and the type of magnetic order in these compounds.        

\section*{Acknowledgments}
This work was financially supported by the Deutsche Forschungsgemeinschaft (DFG) within the Collaborative Research Center SFB 1143 ``Correlated Magnetism – From Frustration to Topology” (project-id 247310070) and the Dresden-Würzburg Cluster of Excellence (EXC 2147) ``ct.qmat - Complexity and Topology in Quantum Matter" (project-id 39085490). K. M. \mbox{acknowledges} the Hallwachs–Röntgen Postdoc Program of ct.qmat for financial support. A. A. acknowledges financial support by the DFG through Grant No.~AL~1771/4-1.

\appendix*
\section{Details on the sample characterization}
\label{sec:sample_details}

\begin{figure}
	\includegraphics[width=0.9\columnwidth]{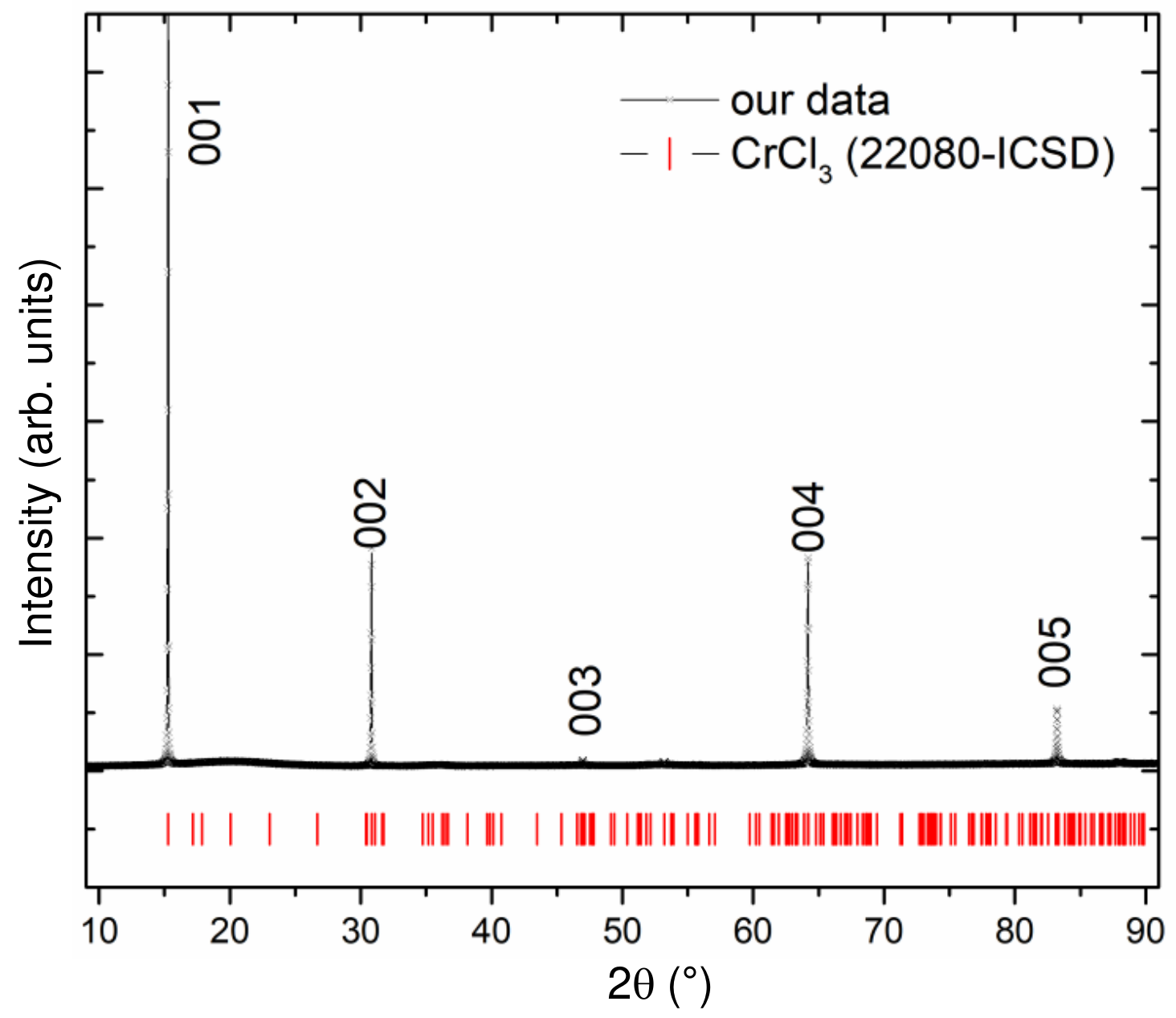}
	\caption{XRD powder pattern of a sample of \CCL. Red ticks display the peak positions of the reference \CCL from the Inorganic Crystal Structure Database  (22080-ICSD, SCXRD, 298\,K). Note that only $(00l)$ peaks are visible due to the strong texturing of the powder sample.}
	\label{fig:x-ray}
\end{figure}

A typical x-ray diffraction (XRD) pattern of a powder sample of \CCL is shown in Fig.~\ref{fig:x-ray}. It was   collected using a PANalytical X’Pert Pro MPD diffractometer with Cu-K$_{\alpha 1}$ radiation ($\lambda = 1.54056$\,\AA) in the Bragg–Brentano geometry at room temperature in the 2$\theta$ range between 10 and 90$^\circ$. The \CCL sample displays strongly preferred orientation leading to only $(00l)$ peaks visible. 

\begin{figure}
	\includegraphics[width=0.9\columnwidth]{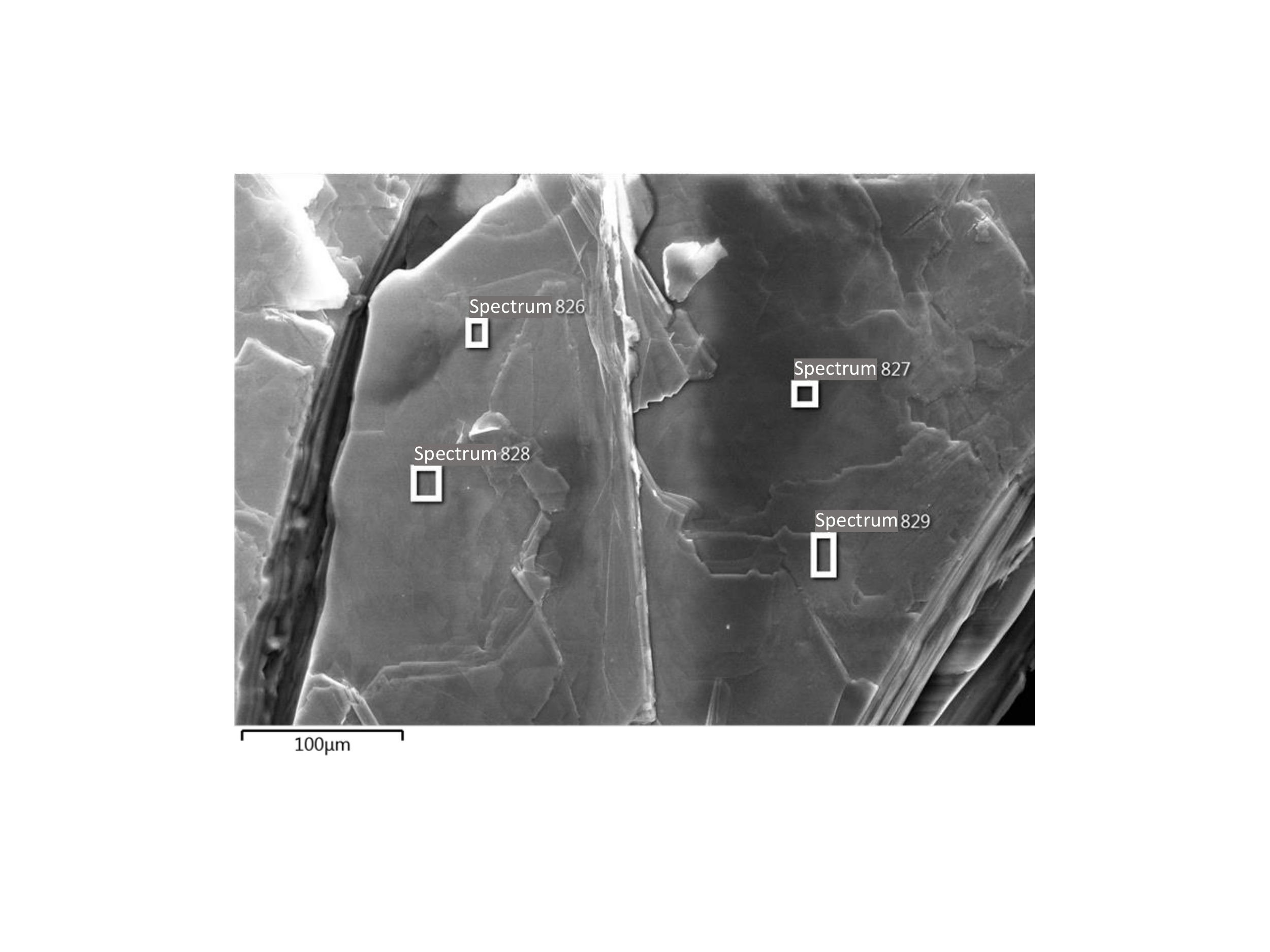}
	\caption{SEM image of \CCL\ crystallites. Rectangles indicate the regions where EDX spectra were collected.}
	\label{fig:SEM_image}
\end{figure}

A typical scanning electron microscopy (SEM) image of \CCL\ crystallites is shown in Fig.~\ref{fig:SEM_image}. It was made with a Hitachi SU8020 microscope equipped with an Oxford Silicon Drift X-MaxN energy dispersive x-ray spectrometer (EDX) at an acceleration voltage of 20\,kV. Results of the EDX analysis at selected areas indicated in Fig.~\ref{fig:SEM_image} are presented in Table~\ref{tab:EDX}. They evidence a very homogeneous, close to the ideal composition Cr\,:\,Cl\,=\,1\,:\,3 of the single crystal.

\begin{table}[!tbh]
	\caption{Results of the analysis of the EDX spectra of the \CCL crystal taken at selected areas indicated in Fig.~\ref{fig:SEM_image}. }
	\setlength\extrarowheight{0.5pt}
	\begin{ruledtabular}
		\begin{tabular}{c c c c c c }
		Spectrum	& 826 & 827 & 828 & 829 & Calculated\\
		 \hline 
			Cl (\%) & 74.77 & 74.69 & 74.49 & 74.87 & 75 \\
			Cr (\%) & 25.23 & 25.31 & 25.51 & 25.13 & 25\\
		\end{tabular}
	\end{ruledtabular}
	\label{tab:EDX}
\end{table}

\bibliography{literature_CrCl3}

\end{document}